\documentclass[aps,prb,reprint,amsmath,amssymb,superscriptaddress]{revtex4-1}
\usepackage{graphicx}
\begin{document}
\title{Hyperpolarized Nanodiamond with Long Spin Relaxation Times}
\author{Ewa Rej$^\dagger$}
\affiliation{ARC Centre of Excellence for Engineered Quantum Systems, School of Physics, University of Sydney, Sydney, NSW 2006, Australia.}
\author{Torsten Gaebel$^\dagger$}
\affiliation{ARC Centre of Excellence for Engineered Quantum Systems, School of Physics, University of Sydney, Sydney, NSW 2006, Australia.}
\author{Thomas Boele}
\affiliation{ARC Centre of Excellence for Engineered Quantum Systems, School of Physics, University of Sydney, Sydney, NSW 2006, Australia.}
\author{David E. J. Waddington}
\affiliation{ARC Centre of Excellence for Engineered Quantum Systems, School of Physics, University of Sydney, Sydney, NSW 2006, Australia.}
\author{David J. Reilly$^*$}
\affiliation{ARC Centre of Excellence for Engineered Quantum Systems, School of Physics, University of Sydney, Sydney, NSW 2006, Australia.}

\begin{abstract}
\bf{The use of hyperpolarized agents in magnetic resonance (MR), such as $^{13}$C-labeled compounds, enables powerful new imaging and detection modalities that stem from a 10,000-fold boost in signal. A major challenge for the future of the hyperpolarizaton technique is the inherently short spin relaxation times, typically  $<$ 60 seconds for $^{13}$C liquid-state compounds, which limit the time that the signal remains boosted. Here, we demonstrate that 1.1$\%$ natural abundance $^{13}$C spins in synthetic nanodiamond (ND) can be hyperpolarized at cryogenic and room temperature without the use of toxic free-radicals, and, owing to their solid-state environment, exhibit relaxation times exceeding 1 hour. Combined with the already established applications of NDs in the life-sciences as inexpensive fluorescent markers and non-cytotoxic substrates for gene and drug delivery, these results extend the theranostic capabilities of nanoscale diamonds into the domain of hyperpolarized MR.}
\end{abstract}
\maketitle

Nanoparticles, having found use in the treatment of cancers \cite{Wang_Thanou,Li}, the study of autoimmune diseases \cite{Getts}, and cardiovascular affections \cite{Gupta}, are currently of interest as theranostic agents needed for the advent of personalised medicine \cite{Mura_Couvreur}. These nanoscale systems are expected to integrate imaging, tracking, and monitoring capabilities with targeted delivery of compounds to tumours, cellular functions and processes, or specific organs. Especially powerful is the modality established by combining high-resolution magnetic resonance imaging (MRI) with nanoparticles that have been hyperpolarized to act as contrast agents, as has been achieved recently using silicon compounds \cite{Cassidy_NN,Cassidy_PRB}. 

Nanodiamonds (ND) are well-suited to act as theranostic platforms, having demonstrated an innate compatibility with biological environments and low toxicity in comparison to other nanoscale structures \cite{Liu_nanotechnology,huang}.  The readily modifiable surface, which is easily functionalized \cite{kruger}, has enabled NDs to be be conjugated to specific molecules \cite{Say_review}, opening a plethora of biomedical applications that include pharmaceutical delivery \cite{huang, chow, chen, zhang} and intracellular tracking \cite{mcguinness} based on the unique optical properties of defects in the diamond lattice \cite{Say_review}. A particular defect, the nitrogen-vacancy (NV) colour centre, has also established a sensitive means of detecting minute magnetic fields on the nanoscale using methods pioneered in controlling quantum devices \cite{maze,balasubramanian,NNReview}. Beyond luminescence-based techniques however, approaches to non-invasively detect and image diamond nanoparticles \textit{in vivo} have to date, been lacking.

Standard MRI modalities (operating at few Tesla magnetic fields) are not well suited to resolving weak concentrations of ND \textit{in vivo} since diamond is a dilute spin system (1.1\% $^{13}$C) and carbon has a small gyromagnetic ratio. This limitation can, in principle, be overcome using hyperpolarization techniques \cite{Keshari_Wilson} which can result in a 10,000-fold boost in signal over that from typical thermal polarization conditions \cite{ardenkjaer,Day}. Hyperpolarized molecular compounds such as [1-$^{13}$C]pyruvate, for example, have recently been used to study tumour metabolism in humans by first transferring electron spin polarization to $^{13}$C nuclei at cryogenic temperatures \cite{nelson}.

In these liquid-state compounds, hyperpolarized $^{13}$C spins typically relax to thermal equilibrium on timescales $T_1$ $<$ 60 seconds \cite{Keshari_Wilson}. In contrast, bulk, high-purity diamond can exhibit $^{13}$C   $T_1$ times of many hours \cite{Hoch_Reynhardt} and recent work using optical techniques to manipulate NV centres \cite{Reimer,Fischer,Alvarez,Pines} has produced significant polarization in large single-crystal samples. The challenge therefore is to maintain these long spin lifetimes even when diamond is produced in nanoparticle form and in sufficient quantities to be of clinical relevance (see Fig. 1a). Addressing this challenge requires a detailed understanding of particle size effects, the structure of internal crystal defects, contaminants, and spin-relaxation channels that arise from the nanoparticle surface \cite{Panich,Shames,casabianca}. Balancing these constraints, the hyperpolarization mechanism also requires the presence of unpaired electrons which, in the case of liquid $^{13}$C compounds are typically added to the agent in the form of organic free-radicals.

In the present work we extend the opportunity for deploying nanodiamond in life-science applications by demonstrating its suitability as a MR marker and contrast agent for MRI. Using electron spin resonance (ESR) we observe that inexpensive commercially available ND, produced via the high pressure high temperature process, surprisingly exhibits a suitable balance of non-toxic paramagnetic centres from defects and surface dangling bonds to allow both hyperpolarization and the preservation of long spin relaxation times. In comparison to previous results on detonation ND \cite{Panich,Shames}, the $^{13}$C relaxation data reported here exhibits a 1,000-fold extension in $T_1$ together with signal enhancements that compare favourably with hyperpolarized $^{13}$C liquid-state compounds. Particle size is found to significantly effect both the relaxation time and amount of achievable hyperpolarization, opening the possibility of using ND to investigate size dependent processes such as the breach of the blood-brain-barrier \cite{BBB}. In addition to showing significant hyperpolarization at $T$ = 4 K, we demonstrate that a sizeable signal enhancement is also possible at liquid nitrogen (77 K) and room temperature using dynamic nuclear polarization (DNP), alleviating the need for expensive liquid helium and potentially enabling new \textit{in vivo} modalities.  Finally, we examine the spin dynamics of the ND core and its surface using hyperpolarized states to resolve new phenomena associated with defects in this versatile material system.\\

\begin{figure}
\includegraphics[scale=1]{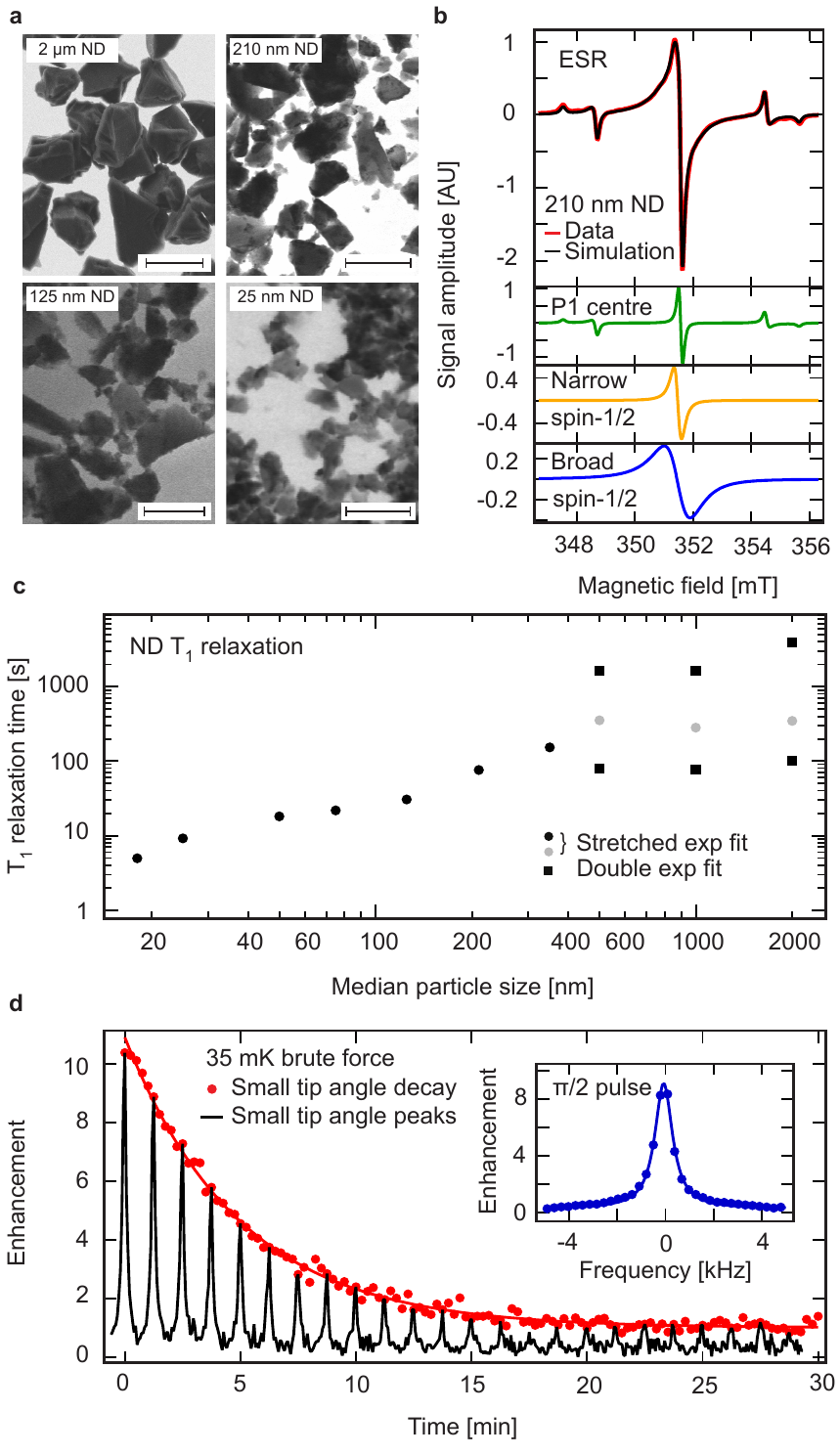}
\caption{\small {\bf (a)} Electron micrographs of various sized NDs used in this work. Top left: scale bar = 2 $\mu$m, top right: scale bar = 400 nm, bottom left: scale bar = 200 nm, bottom right: scale bar = 100 nm.  {\bf(b)} ESR spectrum of 210 nm ND (red). The black line is a simulated spectrum consisting of three components: a narrow spin-1$/$2 component (yellow) a broad spin-1$/$2 component (blue) and a P1-centre component (green) [see Methods Section for details of fitting]. {\bf(c)} Room temperature $^{13}$C relaxation times, $T_1$, as a function of diamond particle size at $B$ = 7 T. Relaxation times were measured using the saturation recovery method with build-up time of magnetization $M$ determined by fitting a stretched exponential M $= M_0(1-exp(-(t/T_1)^\alpha)$ or double exponential. Small NDs exhibit a buildup with $\alpha =$ 2/3 (black circles), with larger NDs better fitted by a stretched exponential with $\alpha =$ 1/2 (grey circles) or double exponential fits with a long and short component shown as black squares. {\bf(d)} Enhanced signal following brute-force hyperpolarization of 2 $\mu$m ND at $T$ = 35 mK and $B$ = 4 T for 3 days in a dilution refrigerator. Following a 40 s transfer in a field of 630 mT, detection is at $B$ = 7 T, via a $\pi/2$-pulse (inset) with decay ($T_1$ $\sim$ 53 min) measured via a sequence of small tip angles (main panel).}
\end{figure}

\begin{figure*}
\includegraphics[scale=1]{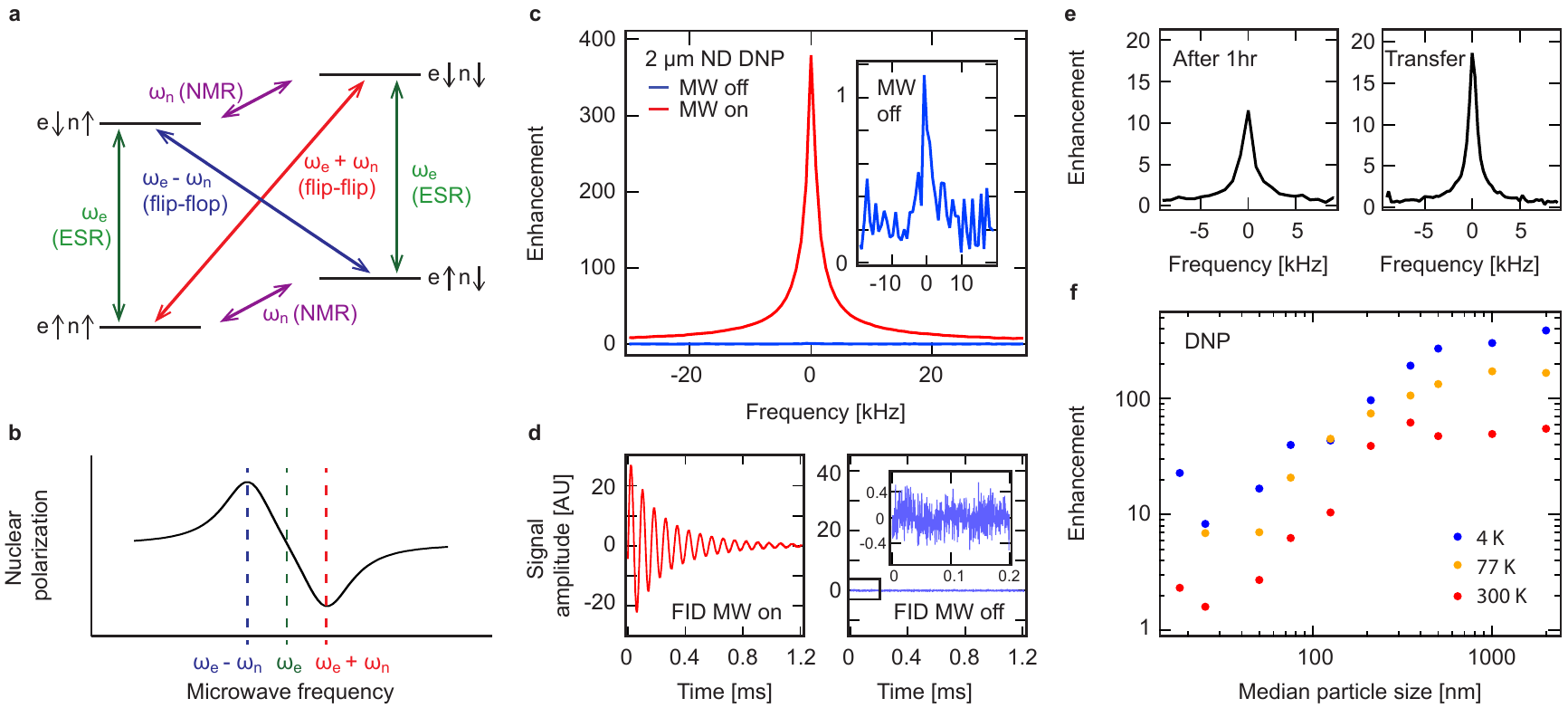}
\caption{\small {\bf (a,b)} DNP via the solid effect is used to hyperpolarize ND. Energy level diagram for a dipolar coupled electron spin-1$/$2 and a nuclear spin-1$/$2  system in a magnetic field. The ESR, NMR, flip-flop and flip-flip transitions are shown. Driven flip-flop transitions (blue) at a frequency $\omega_e - \omega_n$ involve a mutual electron flip and a nuclear flop resulting in a positive nuclear polarization, shown in 2(b). Driven flip-flip transitions (red) result in a negative nuclear polarization.  {\bf (c)} Hyperpolarized signal enhancement of 380 times for 2 $\mu$m ND at $T$ = 4 K. Inset: Zoom of NMR signal taken at thermal equilibrium ($T$ = 4 K) with no microwaves.   {\bf(d)} NMR free induction decay of 2 $\mu$m ND after a $\pi$/2-pulse at $T$ = 4 K. Left: signal after hyperpolarization Right: Thermal polarization signal. {\bf(e)} Left: Signal enhancement of 2 $\mu$m ND that was  polarized for 1 hr and then allowed to decay at field ($B$ = 3 T) for 1 hr. Right: 2 $\mu$m ND signal enhancement after transfer from the polarizer to a $B$ = 7 T magnet for detection. The 2 $\mu$m ND was polarized for 15 min at $T$ = 4 K. The transfer took place in a field of 730 mT and took approx. 15 s. {\bf(f)} Signal enhancement as a function of particle size and temperature $T$ = 4 K (blue), 77 K(yellow) and 300 K (red). The enhancement is given by the hyperpolarized signal divided by the thermal signal at each temperature.}
\end{figure*}

\noindent {\bf ESR spectra and nuclear spin relaxation}\\
Turning to the experimental results [see Methods and Supplementary Section for details], the ESR spectrum and simulation results for a representative sample of 210 nm ND is shown in Fig. 1b. This spectrum, which indicates the predominant types of defects available for use in  hyperpolarization, can be seen to comprise three components that sum to produce the black simulation curve in Fig. 1b. These are a broad spin-1/2 Lorentzian component (blue trace) attributed to carbon dangling bonds near the surface of the ND, a narrow spin-1/2 Lorentzian component (yellow trace) attributed to defects within the diamond lattice, and a component associated with P1-colour centres (green trace) which constitutes a substitutional nitrogen atom with the extra electron hyperfine coupled to the $^{14}$N spin-1 nucleus.  The number of P1-centre impurities, which lead to central ($m_I = 0$) and hyperfine transitions ($m_I =\pm 1$), increases as ND size increases, while the number of spin-1/2  impurities (broad and narrow spectra) decreases as ND size increases [see supplementary material].

These defect sites also provide the primary mechanism for $^{13}$C nuclear spin relaxation in ND. We find that the $T_1$ relaxation time grows with increasing particle size, as shown in Fig. 1c. In determining these $T_1$ times, the spin polarization build-up for smaller diamonds is well described by models \cite{Furman,Blumberg} in which the dipolar interaction of nuclear spins with paramagnetic impurities dominates over nuclear spin diffusion, leading to polarization curves that follow a stretched exponential form [see Methods Section]. Diamond particles with average diameter approaching 1 $\mu$m however, are better characterised by a double exponential in their polarization build-up with time. For 2 $\mu$m diamonds the longer component of the double exponential yields a $T_1$ time of 63 min [see Methods Section and supplementary material]. \\

\noindent{\bf Brute-force hyperpolarization} \\
The simplest method of increasing the MR signal from ND is to first cool the system to low temperatures in a high magnetic field to increase the Boltzmann population difference in the nuclear spins, a process termed `brute-force' polarization. If the NDs are subsequently moved to a different magnetic field and temperature, the spin system can be considered hyperpolarized until it thermalizes on timescale $T_1$. Using the brute force method we hyperpolarize 2 $\mu$m ND at $T$ = 35 mK and $B$ = 4 T in a dilution refrigerator fitted with a rapid sample exchange system that allows fast ($<$ 1 min) transfer of the ND sample to a room temperature $B$ = 7 T magnet for detection. A $\pi$/2 pulse applied immediately after transfer produces a signal (see inset Fig. 1d) that is enhanced by an order-of-magnitude when compared to the signal from 2 $\mu$m ND at thermal equilibrium and $B$ = 7 T. To measure the relaxation time a series of small tip-angles is used to destroy the polarization over 0.5 hours, as indicated by the decaying signal in Fig. 1e. The decay is a combination of the $T_1$ relaxation of spins in the ND ($\sim$ 53 min) and polarization lost from the tipping pulses [see Methods Section for details]. 

\noindent{\bf Hyperpolarization via the solid effect} \\
To achieve even higher polarizations and larger signals, DNP \cite{Abragam} can be used to transfer electron polarization to the $^{13}$C nuclear spins in the diamond \cite{Hoch_Reynhardt}. As described above, the source of these unpaired electrons in ND are paramagnetic centres in the lattice, dipolar coupled to a surrounding nuclear spin bath.  Application of a microwave magnetic field slightly below the electron spin resonance frequency can drive spin flip-flops between nuclear and electron spins associated with centres, leading to a net transfer of spin polarization from the electrons to the nuclei near the impurities in a process known as the solid effect (see Fig. 2a,b). \\

Turning to the main results of our work, we demonstrate that DNP can be used to hyperpolarize commercially available NDs, which as we have shown above, also exhibit long relaxation times. In the case of the largest diameter diamonds  (2 $\mu$m) a $T$ = 4 K signal enhancement of $\sim 400$ is achieved over thermal equilibrium, corresponding to a nuclear polarization of $\sim$ 8 \%, as shown in Fig. 2c,d. Comparing this hyperpolarized signal to the thermal signal at room temperature gives an enhancement of 13,500 $\times$, similar to what has been demonstrated with isotopically labeled $^{13}$C liquid compounds \cite{Keshari_Wilson}. 

It is possible that by hyperpolarizing the nuclear spin system using DNP, new relaxation channels are created that shorten the relaxation time. We test this possibility by first polarizing 2 $\mu$m ND for 1 hr and then allowing it to decay for 1 hr at field. The resulting signal, shown in Fig. 2e, indicates a $T_1$ comparable to measurements performed at thermal equilibrium. As a further demonstration of the potential for hyperpolarized ND, we show in Fig. 2e that the enhanced polarization can be maintained during transfer of the sample from a lower field polarizer to a high field MR detection system [see Methods Section for details].
 
Unlike hyperpolarized molecular compounds, the use of nanoparticles opens a new modality that links MR signal strength (and relaxation time) to particle size. For hyperpolarized ND, we determine a significant size dependence to the signal enhancement, as shown in Fig. 2f. This dependence is most prominent for particle sizes below $\sim$ 300 nm, where the larger rate of spin relaxation competes with the rate at which hyperpolarization from DNP occurs. We suggest that this dependence on the diameter of NDs opens a means of tracking and examining biological functions that are size dependent.

For potential clinical use of hyperpolarized MRI, a major drawback of the technique is the need for liquid helium to cool sample agents during the polarization phase. This drawback is particularly significant for applications that require MR in remote locations \cite{battlefield}. In the case of hyperpolarized ND however, we find that sizeable enhancements are possible at liquid nitrogen temperatures (77 K) (see Fig. 2e), where cryogens are readily available. Extending this idea, Fig. 2e also shows that hyperpolarization is possible at room temperature, doing away with cryogens altogether.\\

\begin{figure}
\includegraphics[scale=1]{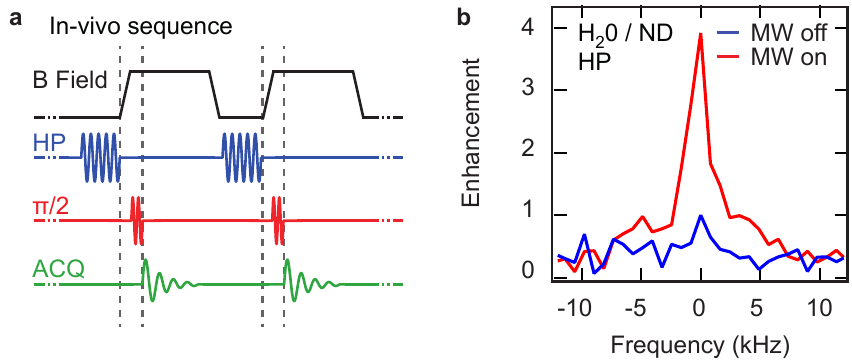}
\caption{\small {\bf(a)} The combination of room temperature signal enhancement and long relaxation times opens the possibility of \textit{in vivo} hyperpolarization using a magnetic field shuttling technique. With the ND agents already administered to the bio-system, hyperpolarization is performed at low-field where microwave heating is reduced. The field is then ramped on a timescale much shorter than $T_1$ to enable imaging and detection of the hyperpolarized ND. Figure shows a repeated sequence with radio frequency $\pi$/2 -pulses and acquisition window (ACQ). {\bf(b)} Hyperpolarization in the presence of water, mimicking \textit{in vivo} conditions. Even at the high microwave frequency of 80 GHz, we observe an enhanced signal (red) of $\sim$ 4 times compared to the thermal polarization (blue). }
\end{figure}

\noindent{\bf Hyperpolarization in the presence of water} \\
Room temperature hyperpolarization, even for modest enhancements, opens the prospect of new \textit{in vivo} modalities when combined with agents exhibiting long $T_1$ times. For instance, the signal from 350 nm ND can be enhanced by a factor of 40 at room temperature and exhibits a $T_1$ of several minutes, sufficiently long to allow the magnetic field to be ramped between polarize and imaging conditions \cite{PEDRI}. \textit{In vivo} hyperpolarization would not only do away with the need for a pre-polarize, and transfer process, it would enable continuous tracking of nanoparticle compounds over exceedingly long times by an interleaved sequence of polarization and detection (see Fig. 3a). The significant barrier to this technique is the heating of water and surrounding tissue during the application of microwaves needed to perform DNP. To test this modality we hyperpolarize a slurry of 125 nm ND and water (200 $\mu$L water with $\sim$ 50 mg ND) at room temperature. Even in the presence of $\sim$ 80 GHz microwaves, we observe a 4-fold enhancement of $^{13}$C MR signal from the diamond with little discernible heating of the water. This enhancement, which corresponds to a halving of the signal relative to the case without water, suggests that such modalities may be possible for small animals or small volumes of tissue polarized at low field (where the DNP frequency is below 1 GHz). \\

\begin{figure}
\includegraphics[scale=0.95]{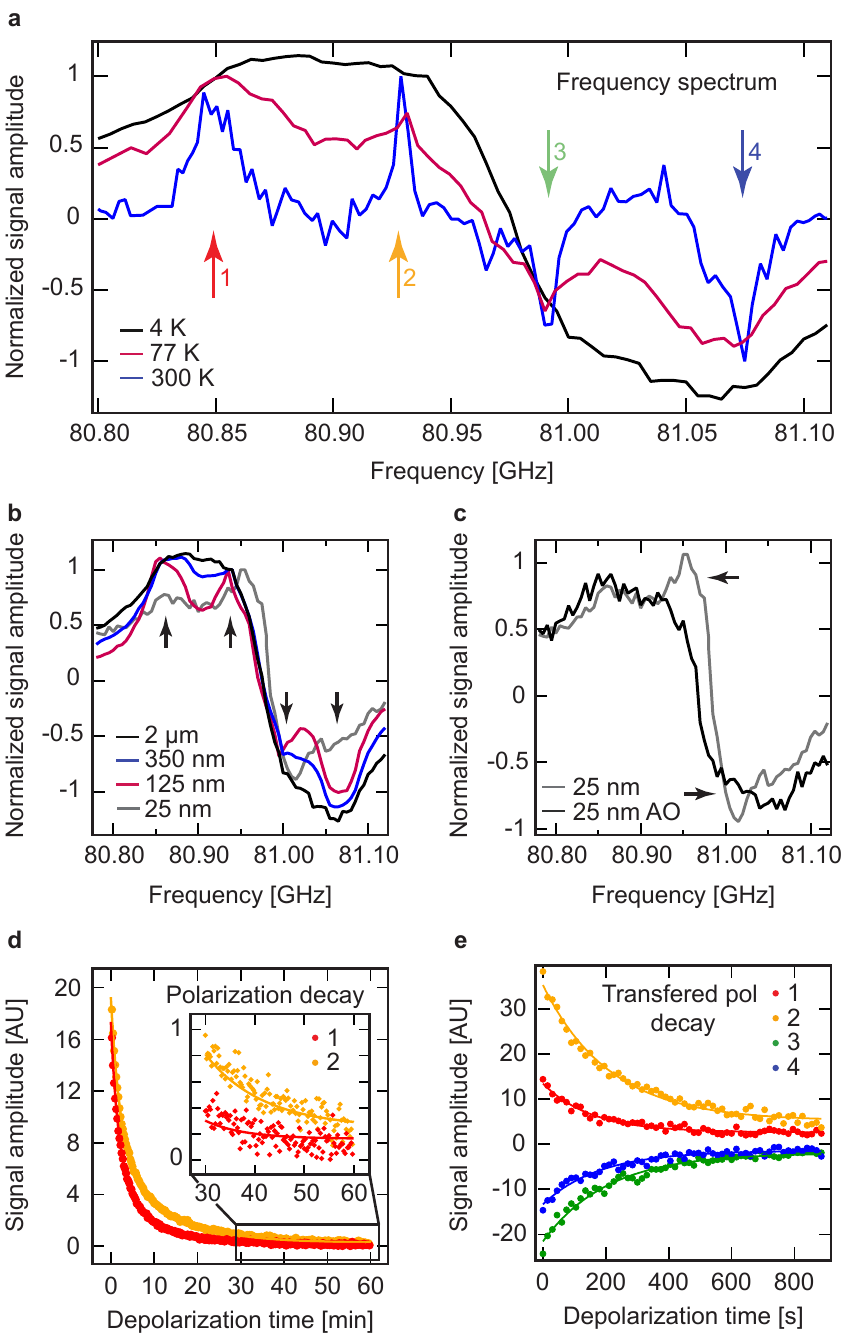}
\caption{\small {\bf (a)} Hyperpolarized signal of 2 $\mu$m ND as a function of polarization frequency at $T$ = 4 K (black), 77 K (purple) and 300 K (blue). The arrows indicate the four frequencies at which polarization build up and decay were examined. The traces were normalized to 1. Note the features broaden as temperature decreases.  {\bf (b)} Normalized hyperpolarized signal of 2 $\mu$m ND (black), 350 nm ND (blue), 125 nm ND (purple), and 25 nm ND (grey) at $T$ = 4 K as a function of microwave frequency. More features are visible as the particle size decreases. {\bf (c)} Comparison of the hyperpolarized signal amplitude as a function of frequency for 25 nm ND and 25 nm air oxidized ND. {\bf (d)} Depolarization of the hyperpolarized signal of 350 nm ND at $T$ = 4 K using a small tip angle pulse sequence ($\theta = 3.5^{\circ}$, $TR = 15\;s$) The ND was polarized for 180 minutes at the frequencies indicated by the arrows in 3(a). We see a difference in the decay times of the red trace ($f$ = 80.870 GHz, $T_1$ = 12 s, 2.1 s) and the yellow trace ($f$ = 80.930 GHz, $T_1$ = 16 s, 2.5 s). For d) and e) the data has been multiplied by $cos(\theta)^{n-1}$ to take into account the polarization lost due to the small tip angle. Fits to the data (solid lines) are double-exponential decay curves. {\bf (e)} Depolarization of the signal from 350 nm ND after hyperpolarization for 15 min at $T$ = 4 K at the four frequencies indicated by arrows in Fig. 3a and transfer to a $T$ = 7 T magnet for detection. The decay was examined using a small tip angle sequence ($\theta = 8^{\circ}$, $TR = 15\;s$). Data is shown in red (80.870\;GHz, $T_1$ = 157\;s), yellow (80.925\;GHz, $T_1$ = 250\;s), green (80.990\;GHz, $T_1$ = 190\;s) and blue (81.050\;GHz, $T_1$ = 157\;s).}
\end{figure}

\noindent{\bf ND Impurity selection and surface modification} \\
The results presented so far are for ND samples readily obtained commercially, without further treatment or surface modification. We now turn to examine the role that surface impurities play in both hyperpolarization and relaxation, noting that there is significant scope to tailor the surface chemistry via passivation and treatment \cite{Surface}. Our approach is to effectively perform ESR spectroscopy at the magnetic field used for DNP ($B$ $\sim$ 3 T), by monitoring the NMR signal enhancement as a function of microwave frequency [see Fig. 4a and supplementary material]. At room temperature we observe enhancement spectra consisting of four peaks that are in agreement with the low-field ESR data shown in Fig. 1b. The position of these peaks correspond to DNP processes at $\omega_e \pm \omega_n$. As the temperature is lowered, these distinct peaks become dipolar broadened. The high field DNP spectra are similarly broadened as the particle size is increased, as shown in Fig. 4b.

Modifying the defects or the types of defects within the nanodiamonds can change the DNP spectra, leading to enhanced polarization and longer relaxation times. We find that burning off the outer layer of ND using air oxidation (AO) processes \cite{Oxidized} removes some of the broad spin-1/2 component associated with impurities near the surface of the ND. This can be seen in Fig. 4c where we compare the hyperpolarization spectra of 25 nm AO ND (black) with standard 25 nm ND (grey). The oxidation process leads to a suppression of the two central lines (2,3) in the spectrum, consistent with removing some of the surface impurities that would otherwise contribute to the signal. 

By adjusting the microwave frequency for DNP, we can select different impurity sites for use in hyperpolarization. Polarization via the P1-centres, for instance, can be selected by irradiating at the frequencies corresponding to the outer peaks (1,4) in Fig. 4a. This is in contrast to irradiating at the inner peaks (2,3) which also comprise both narrow and broad components from spin-1/2 sites (see discussion of Fig. 1b). Surprisingly, we find that hyperpolarization due to microwave driving at the inner peaks takes longer to build up  and is retained longer than when driving at the outer peaks associated with the P1-centres. This behaviour is seen in Fig. 4d [and supplementary material], where for 350 nm ND, we compare the decay of hyperpolarization established by driving at peak 1 or peak 2 in Fig. 4a. These results suggest that nuclear spin diffusion is somewhat suppressed in these systems, since relaxation appears dominated by the particular impurities selected for polarization via the choice of microwave frequency. 

Beyond examining the spin dynamics of ND, these results are of practical interest in optimising conditions for maximum signal enhancement with minimal relaxation. As an example, we compare the signal from 350 nm ND, initially irradiated with microwaves at the four distinct frequencies indicated in Fig. 4a and then subsequently transferred to a $B$ = 7 T system for detection.  Following sample transfer, the relaxation data in Fig. 4e show that a larger polarization is maintained if the nuclei were polarized using the two central spectra peaks (labelled 2 and 3), in comparison to the outer peaks associated with the P1-centres (labelled 1 and 4). 

There is significant scope to further enhance the degree of hyperpolarization in ND. In particular, by adding a microwave capability to our brute force hyperpolarizer at mK temperature (based on a dilution refrigerator), much larger polarization appear possible. Eliminating isolated defects via surface passivation techniques \cite{Surface} will also likely result in longer relaxation times. Another direction is to work with isotopically enriched $^{13}$C NDs to increase the number of spins that contribute to the signal and enhance spin diffusion from polarization sites on the surface to the spins in the core. Finally, we mention the possibility of using NDs for long time storage of nuclear polarization that is transferred to the hydrogen nuclei in an aqueous environment using cross-polarization sequences \cite{CP}. This modality would enable life-science applications in which nanodiamond is tracked and imaged using standard MRI technology.

\section{Methods}
{\bf Nanodiamonds} used in these experiments were purchased from Microdiamant. The diamonds are synthesised using the HPHT technique. The sizes of the NDs are given as a distribution and a median size. We refer to the diamonds by their median size. Measurements were made on MSY 0-0.030, (0-30 nm, median 18), MSY 0-0.05 (0-50 nm, median 25 nm), MSY 0-0.1 (0-100 nm, median 50 nm), MSY 0-0.15 (0-150 nm, median 75 nm), MSY 0-0.25 (0-250 nm, median 125 nm), MSY 0-500 (0-500 nm, median 210 nm), MSY 0.25-0.5 (250 nm-500 nm, median 350 nm), MSY 0.25-0.75 (250 nm-750 nm, median 500 nm), MSY 0.75-1.25 (750-1250 nm, median 1000 nm), and MSY 1.5-2.5 (1500 nm-2500 nm, median 2000 nm). 

SEM measurements were made using a Zeiss Ultra Plus Gemini SEM spectrometer working in transmission mode. Suspensions were made of these NDs in water and a small amount of the suspension was placed upon a TEM grid. The ND size distributions were confirmed (see supplementary material). 

{\bf ESR measurements} were made using a Bruker EMX-plus X-Band ESR Spectrometer. The cavity Q ranged between 5,000 for small ND particles and 10,000 for large ND particles. ESR spectra were taken at 0.25 $\mu$W, (within the linear regime of the saturation curves of the impurities) at a modulation amplitude of 1 Gs and a modulation frequency of 100 kHz. Each of the three components were simulated separately using Easyspin\cite{easyspin} and added together to make the final spectrum. Linewidth, signal amplitude and g-factor were varied. The best fit to the data was performed using a least squares analysis. 

{\bf NMR measurements} to determine the $T_1$ relaxation times at $B$ = 7 T were made using a saturation recovery pulse sequence that involved 64 $\times$ $\pi/2$ saturation pulses to null any initial polarization followed by a varied time for polarization to build up, and then a $\pi/2$ detection pulse. Time domain signals were acquired using a Redstone Tecmag system. Either a stretched exponential (smaller particles) M $= M_0(1-exp(-(t/T_1)^\alpha)$ or a double exponential (larger particles) was fitted to the curve (see supplementary material). Each ND size was measured 3 times and the average of the $T_1$ times is plotted. 

{\bf Decay of brute force hyperpolarization} (red dots) in Fig. 1d was measured using a small tip angle detection sequence ($\theta = 17^{\circ}$, $TR$ = 15 s). The solid red line is a fit to $M = M_0 cos(\theta)^{n-1}e^{-(n-1)TR/T_1}$ resulting in a $T_1$ of 53 min. The decay is a combination of the $T_1$ decay of the particles and the signal lost due to rf induced polarization loss with tip angle $\theta$ where n is the nth pulse and TR is the repetition time. Every fifth peak is shown in black.

{\bf Hyperpolarization measurements} were made at a field of 2.88 T with a Redstone Tecmag system and a in-house constructed NMR probe (design details can be found in Ref \cite{thesis}) inside an Oxford Instruments flow cryostat. The microwave source was a tuneable Gunn Oscillator (80.5 - 81.5 GHz) combined with a power amplifier. Microwaves were coupled to the sample using a waveguide. Polarization transfer measurements were determined using an in-house constructed NMR spectrometer based upon a National Instruments system and an NMR probe at a field of 7 T. Before polarization the signal was saturated with 64 $\times$ $\pi/2$ pulses to null any signal. 
Enhancement measurements were made by hyperpolarizing the ND at f $=$ 80.855 GHz (4K, 77K) and 80.85 GHz (300K) and then detecting the signal with a $\pi/2$ pulse. The hyperpolarized signal was compared to the NMR signal with no microwaves and the same polarization build up time. 
Frequency sweeps: The frequency was swept between 80.78 and 81.12 GHz in discrete steps of 5 MHz and polarization was measured at every point. The 2 $\mu$m ND was polarized for 30 s, 3 min and 3 min at $T$ = 4 K, 77 K, and 300 K respectively. The spectra have been normalized to 1 for easier comparison.  

{\bf Depolarization measurements} at $B$ = 7 T: the ND was polarized for 15 min at four frequencies (f $=$ 80.87, 80.925, 80.99 and 81.05 GHz in successive experiments) and then transferred to a $B$ = 7 T magnet for detection. The transfer was performed in a field of $B \sim$ 0.7 T,  created from rare-earth permanent magnets and took $\sim$ 20 s. A small tip angle pulse sequence with 8 degree pulses was used to detect the signal. Enhancement is compared to ND at thermal equilibrium at $B$ = 7 T. Depolarization measurements at $B$ = 2.88 T: The ND was polarized for 180 minutes at two frequencies and the decay was monitored with small tip angle pulses (every 15 s). Depolarization data was multiplied by $cos(\alpha)^{(n-1)}$ to take into account rf induced depolarization. The resulting data was fitted with a double exponential. 

\section{Acknowledgements}
We thank M. Cassidy for technical contributions in the construction of the hyperpolarization probe and M. Cassidy, and C. Marcus for useful discussions. For SEM measurements the authors acknowledge the facilities and technical assistance of the Australian Centre for Microscopy \& Microanalysis at the University of Sydney. For ESR measurements the authors acknowledge the staff and facilities at the Nuclear Magnetic Resonance Facility at the Mark Wainwright Analytical Centre at the University of New South Wales. This work is supported by the Australian Research Council Centre of Excellence Scheme (Grant No. EQuS CE110001013) and DP1094439. \\

$\dagger$ These authors contributed equally to the work.

* Corresponding author,  e-mail: david.reilly@sydney.edu.au 

\bibliographystyle{naturemag}

\end{document}